\documentclass[12pt,a4]{article}
\usepackage{latexsym}
\usepackage{amsmath}
\usepackage{bm}
\usepackage{graphicx,color}
\usepackage{txfonts}
\newcommand{\bea}{\begin{eqnarray}}
\newcommand{\eea}{\end{eqnarray}}

\title{Superradiance Problem of Bosons and Fermions 
for Rotating Black Holes 
in Bargmann-Wigner Formulation}
\author{Masakatsu KENMOKU\thanks{kenmoku@asuka.phys.nara-wu.ac.jp}\\
Department of Physics, 
Nara Women's University, Nara 630-8506, Japan
}
\date{\empty}

\begin{document}
\maketitle
\abstract{
Bargmann-Wigner equations are formulated to represent bosonic 
fields in terms of fermionic fields in curved spacetime. 
The superradiance phenomena of bosons and fermions 
in rotating black hole spacetime are studied 
in the Bargmann-Wigner formulation. 
As a result of the consistent description between 
scalar bosons and spinor fermions, 
superradiance phenomena of the type of positive frequency $(0<\omega)$ 
and negative momentum near horizon $(p_{H}<0)$ 
are shown not to occur. 
}

\section{Introduction}
\renewcommand{\theequation}{\thesection.\arabic{equation}}
\setcounter{equation}{0}
Recently many black holes have been observed 
and most of all are considered as rotating black holes. 
Many theoretical analysis have been done focused on 
matter behavior around rotating black hole sapcetime.  
The framework of the general relativistic theory 
is a reliable approach to the matter perturbation 
outside the black hole horizon in four dimensions.

As one of standard radiation problems of matter fields around 
rotating black holes, the outstanding 
super-radiance problem may occur in which 
the reflected intensity becomes stronger than the incident intensity. 
The successive occurrences of superradiance phenomena will 
cause the serious problem of the instability of black holes 
\cite{Press1972,Chandrasekhar1983,Cardoso2004,Kodama2008}.

Under Kerr background spacetime, 
Klein-Gordon, Dirac, Maxwell, Rarita-Shwinger 
and Einstein equations for massless fields are known to reduce to the 
separable one component field equations, 
which are called as the Teukolsky equations \cite{Teukolsky1972-1974}. 
The analytic perturbative solutions of the Teukolsky equations show that 
the reflected intensity can become stronger than the incident 
intensity for bosons and 
the super-radiance occurs for strongly rotating black hole sapcetime. 
The situation is summarized according to the work by S. Mano and
E. Takasugi \cite{Takasugi1997} as follows:
\[
\mid A_{s}^{(\rm inc)}\mid^{2}
=\frac{2\omega^{2s}}{\mid C_{s}\mid^2}\mid A_{s}^{(\rm ref)}\mid^{2}
+\delta_{s}\mid A_{s}^{(\rm trans)}\mid^{2}\ , 
\]
where $A_{s}^{(\rm inc)}\, , A_{s}^{(\rm ref)}$ and $A_{s}^{(\rm trans)}$ 
denote the amplitudes of incident, reflected and transmitted 
waves respectively with spin $s\,(=0,1/2, \cdots , 2)$ and frequency $\omega$. 
The Starobinsky constants are denoted by $C_{s}$ and coefficients $\delta_{s}$ 
are defined:
\[
\delta_{s}=-i\exp{(i\pi s)}\,
(\frac{\omega}{\sqrt{M^2-a^2}})^{2s-1}\,
\frac{\Gamma(1-s+2i\epsilon_{+})}{\Gamma(s+2i\epsilon_{+})}\ , 
\]
where $M\, , a$ denote the mass and angular momentum of Kerr black holes. 
The constant $\epsilon_{+}$ is defined as
\[
\epsilon_{+}=(1-(a/M))^{-1/2}\,r_{H}\,p_{H}\ ,  
\]
where $r_{H}$ and $p_{H}$ 
denote the radius of the event horizon of black holes and momentum 
of fields near horizon, which is defined by
\[
r_{H}=M+\sqrt{M^2-a^2}\ \ , \ \ 
p_{H}=\omega-m\Omega_{H} \ , 
\] 
with the angular velocity of black holes $\Omega_{H}=a/(r_{H}^2+a^2)$. 
The coefficients $\delta_{s}$ determine 
whether the super-radiance occurs or not according to their sign. 
Those for fermions $\delta_{1/2}\, , \delta_{3/2}$ are shown to be 
positive definite values 
while those for bosons 
\[
\delta_{0}\ , \delta_{1}\ ,\delta_{2}\  \propto \ p_{H} \ , 
\]
can be negative for large angular momentum of black holes. 
That is, the super-radiance occurs for bosons in case of $p_{H}<0$, 
but not for fermions. 

We list up some special features in the super-radiance phenomena 
of rotation black holes as follows: 
\begin{itemize}
\item[1)] 
The super-radiance may occur for bosonic fields 
to the strongly rotating black holes 
(or light mass black holes) 
but not to the weakly rotating black holes 
(or heavy mass black holes)
in four dimensional spacetime \cite{Press1972,Misner1972,
Detweiler1980,Mukohyama2000}. 
\item[2)]
The super-radiance do not occur for fermionic fields 
to any rotating black holes in four dimensional spacetime 
for massive fields \cite{Maeda1976} 
as well as massless fields \cite{Unruh1974}.
\item[3)]
In the three dimensional case,  
the super-radiance phenomena have been shown not to occur 
\cite{Kenmoku2008,Kenmoku2008-2}. 
\end{itemize}

The purpose of this paper is to resolve the puzzle 
in super-radiance phenomena for fermions and bosons 
using the Bargmann-Wigner formulation, which connect 
fermions and bosons through the Bargmann-Wigner equations. 
As a result of our study, we obtain the result that 
the momentum near horizon cannot be negative, 
therefore the superradiance for scalar bosons 
$0<\omega$ and $p_{H}=\omega-m\Omega_{H}<0$ 
does not occur, which is  
in coincident with those for fermions. 

The organization of this paper is as follows.
In section 2, the Bargmann-Wigner equations in flat spacetime 
are reviewed briefly 
in the sake of the following study. 
In section 3, the Bargmann-Wigner equations are extended 
and formulated in general curved spacetime in the case of 
bi-spinor fields. 
In section 4, the Bargmann-Wigner formulation is applied 
to the fermion and boson puzzle in super-radiance phenomena 
in Kerr geometry. 
Summary is given in the final section. 

\section{Bargmann-Wigner formulation in flat spacetime}
\setcounter{equation}{0}
In this section, we review briefly the original Bargmann-Wigner equations 
in flat spacetime. 
An explicit realization is the case for spin 1 states \cite{Bargmann1948,Lurie}.

\subsection{Original Bargmann-Wigner equations in flat spacetime}
We start to short review of the original Bargmann-Wigner equations, 
which are aimed to get a general system of relativistic wave equations 
for higher spin states 
\begin{eqnarray}
(\gamma^{\lambda} {\partial}_{\lambda}+\mu)_{\alpha \alpha^{'}}
{\Psi(x)}^{(\rm BW)}_{\alpha^{'} \beta \dots}&=&0  \nonumber  \\ 
(\gamma^{\lambda}{\partial}_{\lambda}+\mu)_{\beta \beta^{'}}
{\Psi(x)}^{(\rm BW)}_{\alpha \beta^{'} \dots}&=&0 \nonumber \\ 
\cdots \ , 
\end{eqnarray}
where $\mu$ denotes the mass of particles and 
multi-spinor fields $\Psi^{(\rm BW)}$ are assumed top be symmetric 
with respect to their spinor suffices:
$
\Psi(x)^{(\rm BW)}_{\alpha \beta \dots}
=\Psi(x)^{(\rm BW)}_{\beta \alpha \dots}\ .
$

\subsection{Bargmann-Wigner equations for spin 1 states} 
 
For spin 1 states, 
the Bargmann-Wigner equations reduce to two equations 
as follows
\begin{eqnarray}
(\gamma^{\lambda}{\partial}_{\lambda}+\mu)\,{\Psi(x)}^{(\rm BW)}&=&0 \, \label{BW01} \\ 
{\Psi(x)}^{(\rm BW)}\,
({\overleftarrow{\partial}_{\lambda}\gamma^{{\lambda} T} }+\mu)
&=&0\ ,  \label{BW02}
\end{eqnarray}
where $T$ denotes the transpose operation. 
We write the bi-spinor field in the boson expansion form:
\begin{eqnarray}
\Psi(x)^{(\rm BW)}= \sqrt{\mu} \gamma^{\lambda}CA_{\lambda}(x)
+\frac{1}{2\sqrt{\mu}}\Sigma^{\lambda\tau}CF_{\lambda\tau}(x) \ ,
\end{eqnarray}
where $A_{\lambda}$ and $F_{\lambda\tau}$ are vector and anti-symmetric 
second-rank tensor fields and $\Sigma^{\lambda\tau}$ and $C$ 
are the spin tensor and 
the charge conjugation matrix respectively. 
We apply the Bargmann-Wigner equations and get the coupled equations 
for bosons: 
\begin{eqnarray}
F_{\lambda\tau}(x)&=&\partial_{\lambda}A_{\tau}(x)
-\partial_{\tau}A_{\lambda}(x) \ , \label{Beq01}\\
\partial^{\lambda}F_{\lambda\tau}&=&\mu^2A_{\tau}(x) \ , \label{Beq02} 
\end{eqnarray}
which derive the massive vector field equations. 
The Bargmann-Wigner system shows the consistency 
between fermionic expression in eqs.(\ref{BW01})-(\ref{BW02}) and bosonic expression 
in eqs.(\ref{Beq01})-(\ref{Beq02}), which represent the same objects from two sides.

In the following study, 
it is convenient to introduce to define the modified second rank 
Bargmann-Wigner field: 
\begin{eqnarray}
\Psi(x)=\Psi(x)^{(\rm BW)}C^{-1}\gamma_{5}
\end{eqnarray}
to avoid the transpose and charge conjugation operations as
\begin{eqnarray}
(\gamma^{\lambda}\partial_{\lambda}+\mu)\Psi(x)&=&0 \ , \\ 
\Psi(x)
(\overleftarrow{\partial}_{\lambda}\gamma^{\lambda}+\mu)&=&0 \ .
\end{eqnarray}

\section{Bargmenn-Wigner formulation in curved spacetime}
\label{secBW}
\setcounter{equation}{0}

Next we develop the Bargmann-Wigner system, establish the Lagrangian formalism and derive the current conservation 
law in curved spacetime. 

\subsection{Notation and definition in local Minkowski and curved spacetime}

In curved spacetime, the gamma matrices and the algebra are defined 
in local Minkowski space with the Latin letters for suffix notation 
\cite{Brill1957,Weinberg}
\footnote{See appendix for the explicit expression of gamma matrices.}:  
\begin{eqnarray}
\{\gamma^{i}, \ \gamma^{j}\}=2\ \eta^{ij}\ , \
\eta^{ij}={\rm diag}(-1,1,1,1)\ .
\end{eqnarray}
The gamma matrices and the algebra in curved spacetime are obtained from those 
in local Minkowski spacetime 
with the Greek letters for suffix notation: 
\begin{eqnarray}
\gamma^{\mu}&:=&b_{i}^{\mu}\gamma^{i}\ , \\
\{\gamma^{\mu}, \ \gamma^{\nu}\}&=&2\ g^{\mu\nu}\ , \
g^{\mu\nu}=b^{i\ \mu}\ b_{i}\ ^{\nu} \ ,
\end{eqnarray}
where $b_{i}^{\mu}$ denote the vierbein, 
which connect the physical quantities in flat Minkowski spacetime 
with those in curved spacetime, and $g^{\mu\nu}$ denote 
the metric tensor in curved spacetime.  


The spinor fields $\psi(x)$ are introduced to transform as spinors 
under local Lorentz transformations in local Minokowski sapcetime 
and scalars under general coordinate transformations in curved spacetime: 
\begin{eqnarray}
\psi(x) \rightarrow D(\Lambda(x))\psi(x) \ ,
\end{eqnarray}
where $D(\Lambda)$ is the spinor representation of the homogeneous Lorentz 
group. 
A covariant derivative for spinors is introduced as
\begin{eqnarray}
\mathcal{D}_{\mu}\psi=(\partial_{\mu}+\Omega_{\mu})\psi \ , 
\end{eqnarray}
which is defined to transform under local Lorentz transformations 
like $\psi$ itself: 
\begin{eqnarray}
\mathcal{D}_{\mu}\psi(x) \rightarrow D(\Lambda(x))\mathcal{D}_{\mu}\psi(x)\ .\end{eqnarray} 
The connection matrix can be written in the form 
\begin{eqnarray}
\Omega_{\mu}= \frac{1}{2}\omega^{ij}_{, \mu}\Sigma_{ij} \ , \label{Omega}
\end{eqnarray}
where $\omega^{ij}_{, \mu}$ is the spin connection and 
$\Sigma_{ij}$ are the spin matrices representing 
the generators of homogeneous Lorentz group 
and for spinors they are written as 
\begin{eqnarray}
\Sigma_{ij}:=\frac{1}{2}(\gamma_{i}\gamma_{j}-\gamma_{\j}\gamma_{\i})\ .
\end{eqnarray}
The covariant Dirac equation in curved spacetime is derived 
using the defined notations as
\begin{eqnarray}
(\gamma^{\mu}(\partial_{\mu}+\Omega_{\mu})+\mu)\psi(x)=0\ ,
\end{eqnarray}
where $\mu$ is the mass parameter. 

For vector fields $A_{\nu}(x)$ transforming like vectors under 
general coordinate transformations, 
a covariant derivative is defined as
\begin{eqnarray}
\nabla_{\mu}A_{\nu}
=\partial_{\mu}A_{\nu}-\Gamma^{\lambda}_{\nu\mu}A_{\lambda}\ , 
\end{eqnarray}
where $\Gamma^{\lambda}_{\ \ \nu\mu}$ denote the affine connection. 

The covariant derivatives for vierbeins with suffix of 
general curved and local Minkowski spacetime 
are defined: 
\begin{eqnarray}
\mathcal{D}_{\mu}b^{i}_{\ \nu}
=\partial_{\mu}b^{i}_{\ \nu}+\omega^{i}_{\ j, \ \mu}b^{j}_{\ \nu}
-\Gamma^{\lambda}_{\mu \nu}b^{i}_{\lambda} \ .
\end{eqnarray}
In order to determine the geometry the vierbein condition 
is imposed: 
\begin{eqnarray}
\mathcal{D}_{\mu}b^{i}_{\ \nu}=0 \ , \label{condition}
\end{eqnarray}
which leads the explicit form of the spin connection 
\begin{eqnarray}
\omega^{ij}_{,\, \mu}&=&g^{\nu\lambda}b^{i}_{\nu}
(\partial_{\mu}b^{j}_{\lambda}-\Gamma^{\tau}_{\lambda\mu}b^{j}_{\tau}) \ .  
\label{omega}
\end{eqnarray}

\subsection{Bargmann-Wigner equations for spin 0 and 1 states 
in curved spacetime}

We consider the bi-spinor Bargmann-Wigner fields $\Psi(x)$ as 
spin 0 and 1 bosonic states taking off the symmetry restriction 
with respect to spinor suffices. 
The bi-spinor fields transform under local Lorentz transformations 
like spinors from lefthand side and righthand side as 
\begin{eqnarray}
\Psi(x) \rightarrow D(\Lambda(x))\Psi(x) D^{-1}(\Lambda(x)) \ ,
\end{eqnarray}
which define the covariant derivatives for bi-spinors as 
\begin{eqnarray}
\mathcal{D}_{\mu}\Psi(x):=
\partial_{\mu}\Psi+\Omega_{\mu}\Psi-\Psi\Omega_{\mu}\ .
\end{eqnarray}
Then the covariant field equations for bi-spinors are derived
\begin{eqnarray}
(\gamma^{\mu}\mathcal{D}_{\mu}+\mu)\Psi(x) &=&0 \ , \label{BW1}\\
\Psi(x)(\overleftarrow{\mathcal{D}}_{\mu}\gamma^{\mu}+\mu)&=&0 \ .\label{BW2}
\end{eqnarray}
We expand the bi-spinor field in a set of bosons as 
\begin{eqnarray}
\Psi(x)={\sqrt{\mu}}
(S(x)+\gamma_{5}P(x)-\gamma^{\mu}V_{\mu}(x)
+\gamma_{5}\gamma^{\mu}A_{\mu}(x))
+\frac{1}{2{\sqrt{\mu}}}\gamma_{5}\Sigma^{\mu\nu}F_{\mu\nu}(x) \ , \label{BsBo}
\end{eqnarray}
where $S, P, V_{\mu}, A_{\mu}$ and $F_{\mu\nu}$ denote the scalar, 
pseudoscalar, vector and tensor fields respectively. 
We apply the Bargmann-Wigner equations 
in curved spacetime (\ref{BW1}) and (\ref{BW2}). 
Adding these equations, we find the set of relations among bosons 
\begin{eqnarray}
\mu S(x)&=&\nabla^{\lambda}V_{\lambda}(x)\ \label{SVrelation}\\
\mu V_{\lambda}(x)&=&\partial_{\lambda}S(x) \label{VSrelation}\\
P(x)&=&0 \\
\mu^2A_{\lambda}(x)&=&\nabla^{\tau}F_{\tau\lambda}(x) \\
F_{\lambda\tau}(x)&=&\nabla_{\lambda}A_{\tau}(x)-\nabla_{\tau}A_{\lambda}(x) 
\label{FArelation} \ .
\end{eqnarray}
For independent spin 0 and 1 fields, the Klein-Gordon type field equations 
in curved spacetime are obtained as in the flat spacetime:
\begin{eqnarray}
(\nabla^{\mu}\partial_{\mu}-{\mu}^2)S(x)&=&0 \ ,\label{S1}\\
\nabla^{\mu}(\nabla_{\mu}A_{\nu}-\nabla_{\nu}A_{\mu})
-{\mu}^2A_{\nu}(x)&=&0 \ . 
\label{A1}
\end{eqnarray}

The difference of the Bargmann-Wigner equations (\ref{BW1})-(\ref{BW2}) 
is calculated to be 
\begin{eqnarray}
&&(\gamma^{\nu}\mathcal{D}_{\nu}+\mu)\Psi
-\Psi({\overleftarrow{\mathcal{D}}}_{\nu}\gamma^{\nu}+\mu)\ \nonumber \\
&=&2\sqrt{\mu}(\gamma_{5}\gamma^{\nu}\partial_{\nu}P
+\Sigma^{\nu\,\lambda}\nabla_{\nu}V_{\lambda}
-\gamma_{5}\nabla^{\nu}A_{\nu})
-\frac{1}{\sqrt{\mu}}\gamma_{5}
\Sigma^{\nu\lambda\tau}\nabla_{\nu}F_{\lambda\,\tau}
\ , \label{difference}
\end{eqnarray}
where $\Sigma^{\nu\lambda\tau}$ denotes the 
third rand completely anti-symmetric tensor of gamma matrices.
For each term in eq.(\ref{difference}), we use bosonic field relations 
in eqs.(\ref{SVrelation})-(\ref{FArelation}) and obtain
\begin{eqnarray}
\Sigma^{\nu\,\lambda}\nabla_{\nu}V_{\lambda}
&=&\frac{1}{2\mu}\Sigma^{\nu\lambda}
(\Gamma^{\tau}_{\nu\lambda}-\Gamma^{\tau}_{\lambda\nu})\partial_{\tau}S\ , \\\nabla^{\nu}A_{\nu}
&=&\frac{1}{2\mu^2}g^{\nu\rho}g^{\lambda\sigma}
(R^{\tau}_{\rho\ , \lambda\nu}F_{\tau\sigma}
+R^{\tau}_{\sigma\ , \lambda\nu}F_{\tau\rho}
+(\Gamma^{\tau}_{\nu\lambda}-\Gamma^{\tau}_{\lambda\nu})
\nabla_{\tau}F_{\rho\sigma})\ , \\
\Sigma^{\nu\lambda\tau}\nabla_{\nu}F_{\lambda\,\tau}
&=&\Sigma^{\nu\lambda\tau}(R^{\rho}_{\tau\ , \nu\lambda}A_{\rho}
+(\Gamma^{\rho}_{\nu\lambda}-\Gamma^{\rho}_{\lambda\nu})\nabla_{\rho}A_{\tau})
\ , 
\end{eqnarray} 
where $R^{\tau}_{\rho\ , \lambda\nu}$ denotes 
the Riemann-Christoffel curvature tensor. 
In order to satisfy the difference of the Bargmann-Wigner equations, 
we require the following conditions: 
\begin{itemize}
\item[(C1)]The torsion of the background geometry should be zero:
$\Gamma^{\rho}_{\nu\lambda}-\Gamma^{\rho}_{\lambda\nu}=0$, 
for non-zero scalar field S(x). 
\item[(C2)]Axial-vector field should vanish: 
$A_{\nu}(x)=0$ and therefore $F_{\nu\lambda}(x)=0$, 
for general curved geometry of non-zero curvature.  
\end{itemize}
We set the above two conditions (C1) and (C2) in the following of this paper 
\footnote{For the bi-spinor field we impose the anti-symmetric spinor 
suffix: $\Psi^{(\rm BW)}_{\alpha\beta}(x)=-\Psi^{(\rm BW)}_{\beta\alpha}(x)$  
instead of the original Bargmann-Wigner field in which spinor suffix is 
completely symmetric.}.

Here we stress that the solution relations between 
fermionic states and bosonic states. 
If the bosonic solutions $S(x)$ and $A_{\mu}(x)$ 
of equations (\ref{S1}) and (\ref{A1}) are obtained, 
they are automatically the bi-spinor field solutions $\Psi(x)$ 
of (\ref{BW1}) and (\ref{BW2}).
They are the same physical objects from different sides. 
Furthermore the bi-spinor fields are transformed to 
a set of four Dirac spinors $\psi^{(i)}\ , \ (i=1\sim 4)$ as 
\begin{eqnarray}
\Psi(x)U(x)
=(\psi^{(1)}(x),\psi^{(2)}(x),\psi^{(3)}(x),\psi^{(4)}(x))\ ,
\end{eqnarray}
where the local transformation $U(x)$ is defined by 
$U^{-1}\partial_{\mu}U=\Omega_{\mu}$.  
Then each spinor field satisfies the Dirac equation: 
\begin{eqnarray}
(\gamma^{\mu}(\partial_{\mu}+\Omega_{\mu})-\mu)\psi(x)^{(i)}(x)=0\ \ \ 
(i=1\sim 4).
\end{eqnarray}
Therefore the existence of stable scalar bosonic solutions is 
directly related to the existence of stable spinor fermionic states. 
Both of bosonic and fermionic view points are the same from different 
sides and the inverse statement is also correct. 

\subsection{Lagrangian formalism and conserved current}

In order to establish the Bargmann-Wigner formulation in curved 
spacetime, we introduce the Lagrangian formalism. 
The Lagrangian density for the bi-spinor fields is proposed to be
\begin{eqnarray}
\mathcal{L}
=-\frac{1}{8}{\rm Tr} \{ \bar{\Psi}(x)
(\gamma^{\mu}\mathcal{D}_{\mu}+\mu)\Psi(x)
+{\Psi}(x)(\overleftarrow{\mathcal{D}}_{\mu}\gamma^{\mu}+\mu)\bar{\Psi}(x)\} 
\ ,
\end{eqnarray}
where the adjoint spinor is defined as 
\begin{eqnarray}
\bar{\Psi}=(-i\gamma_{0}){\Psi}^{\dagger}(-i\gamma_{0}) \ .
\end{eqnarray}
From this Lagrangian density, the sum of the Bargmann-Wgner equations 
(\ref{BW1}) and (\ref{BW2}) are obtained. 
The Lagrangian density in the bi-spinor form reproduces 
the correct Lagrangian density in the bosonic form 
expressing the independent bosonic fields $S(x)$ and $A_{\mu}(x)$ as 
\begin{eqnarray}
\mathcal{L}
&=&-\partial^{\mu}S^{\dagger}(x)\partial_{\mu}S(x)
-\mu^2S^{\dagger}(x)S(x)  \ ,
\end{eqnarray}
which leads the correct equation of motion. 

From the Lagrangian, the conserved current is derived 
as the Noether current for the phase transformation to bi-spinor 
fields: 
\begin{eqnarray}
\Psi(x)\rightarrow \exp{(i\alpha(x))}\Psi(x) \ .
\end{eqnarray}
The invariance of the Lagrangian under the phase transformation leads 
the conserved current in the bi-spinor expression 
\begin{eqnarray}
J_{\mu}=-\frac{\delta \mathcal{L}}{\delta\partial_{\mu}\alpha(x)}
=\frac{i}{4}{\rm Tr}\ \bar{\Psi}\gamma_{\mu}\Psi \ , \label{currentBs}
\end{eqnarray}
and in the boson expression 
\begin{eqnarray}
J_{\mu}=-i(S^{\dagger}\partial_{\mu}S-\partial_{\mu}S^{\dagger}S 
)\ . \label{currentBo}
\end{eqnarray}
The conservation of the current can be shown using the field equations directly \begin{eqnarray}
\nabla^{\mu}J_{\mu}=0 \ .
\end{eqnarray} 

\section{Application to super-radiance puzzle 
between bosons and fermions in rotating black holes}
\setcounter{equation}{0}
We apply the Bargmann-Winger formulation, which was established in 
section \ref{secBW}, to the super-radiance puzzle 
between bosons and fermions. 
For this purpose, we first study a simple application to free spherical 
waves. 
Next we apply this result to the asymptotic infinity 
region of Kerr black hole geometry 
and derive the current relation between bosons and fermion near the 
event horizon to show no occurrence of the bosonic super-radiance.

\subsection{Case of free spherical waves}

To understand the relation between bosonic and fermionic description 
explicitly, 
we consider free spherical waves as a simple application in flat spacetime. 
For a scalar boson example, 
the solution of Klein-Gordon equation is expressed 
by a spherical wave in the polar coordinate as 
\begin{eqnarray}
S^{(0)}(x)=Y^{m}_{\ell}(\theta\, ,\phi)R_{\ell}(r)\exp{(-i\omega t)}
\ ,
\end{eqnarray}
where $Y^{m}_{\ell}\, ,R_{\ell}$ and $\omega$ denote 
the spherical harmonics, the radial wave function and the frequency 
respectively. The normalization factor for boson is assumed to be 
included in the radial function. 

For this bosonic solution the corresponding bi-spinor solution is obtained 
through equations (\ref{BsBo}) and (\ref{VSrelation}) in the matrix form as 
\begin{eqnarray}
\Psi^{(0)}(x)
&=&(\mu-\gamma^{\lambda}\partial_{\lambda})\frac{1}{\sqrt{\mu}}S^{(0)}(x)\ , \nonumber\\
&=&\frac{1}{\sqrt{\mu}}
\left(
\begin{array}{cc}
\mu+\omega                 & i\boldmath{\sigma}\cdot \nabla \\
-i\boldmath{\sigma}\cdot \nabla & \mu-\omega 
\end{array}
\right) Y^{m}_{\ell}(\theta\,\phi)R_{\ell}(r)\exp{(-i\omega t)} . 
\label{p0}
\end{eqnarray}
We write this as a set of four spinors as 
\begin{eqnarray}
\Psi^{(0)}(x)=\sqrt{\frac{\omega+\mu}{\mu}}(\psi^{(1)}(x),\psi^{(2)}(x),
\psi^{(3)}(x),\psi^{(4)}(x))\ ,
\end{eqnarray}
where $\psi^{(1)},\ \psi^{(2)} $ are spinors of 
the positive energy with spin up and down respectively and 
$\psi^{(3)}, \ \psi^{(4)}$ are those of the negative energy. 
The angular wave functions $Y^m_{\ell}$ 
can be written by the combination of normalized spin-angular functions 
$\mathcal{Y}$ as
\begin{eqnarray}
Y^{m}_{\ell}(\theta\, ,\phi)
\left(
\begin{array}{c}
1\\0
\end{array}
\right)
=\sqrt{\frac{\ell+m+1}{2\ell+1}}\mathcal{Y}^{j_{3}}_{j,\ell}(\theta\, ,\phi) 
-\sqrt{\frac{\ell-m}{2\ell+1}}\mathcal{Y}^{j_{3}}_{j',\ell}(\theta\, ,\phi) , 
\end{eqnarray}
with total angular momenta $j=\ell+1/2, j'=\ell-1/2$ and 
their azimuthal compoment $j_{3}=m+1/2$. 
Corresponding to the recombination, 
the positive energy and spin up spinor can be written as \cite{Sakurai1967}
\begin{eqnarray}
\psi^{(1)}=\sqrt{\frac{\ell+m+1}{2\ell+1}}\psi^{(+)}(x)
-\sqrt{\frac{\ell-m}{2\ell+1}}\psi^{(-)}(x)\ ,
\end{eqnarray}
where the spin parallel and antiparallel spinors $\psi^{({\pm})}$ are denoted 
\begin{eqnarray}
\psi^{({+})}=\frac{1}{r}
\left(
\begin{array}{c}
F(r)\mathcal{Y}^{j_{3}}_{j,\ell}\\
iG^{(+)}(r)\mathcal{Y}^{j_{3}}_{j,\ell+1} 
\end{array}
\right)\ \ \ , \ \ \ 
\psi^{({-})}=\frac{1}{r}
\left(
\begin{array}{c}
F(r)\mathcal{Y}^{j_{3}}_{j',\ell}\\
iG^{(-)}(r)\mathcal{Y}^{j_{3}}_{j',\ell-1} 
\end{array}
\right)\ .
\end{eqnarray}
The fermion radial functions are obtained by the explicit calculation of 
the operation $-i\boldmath{\sigma}\cdot\nabla$ in equation (\ref{p0}) as
\begin{eqnarray}
F(r)&=&\sqrt{\omega+\mu}\,r\,R_{\ell}(r)\ , \nonumber\\
G^{(+)}(r)&=&\frac{(\partial_{r}-({\ell}+1)/{r})F(r)}{\omega+\mu}\ , \ 
G^{(-)}(r)=\frac{(\partial_{r}+\ell/r)F(r)}{\omega+\mu}\ , 
\end{eqnarray}
where the normalization factor  for fermion relative to the boson 
$\sqrt{\omega+\mu}$ is included in the radial function $F(r)$. 
Fermion normalization factor $\sqrt{\omega+\mu}$ is included relative to 
bosonic one. 
Other spin states $\psi^{(2\sim 4)}$ can be obtained similarly.

We have established the relation between scalar wave function $S$ 
and spinor $\psi^{(1)}$ or $\psi^{(\pm)}$ using the bi-spinor field 
$\Psi$ explicitly in the spherical symmetric case. 
Next we consider the radial component of the conserved current given by 
\begin{eqnarray}
J^{(0)}_{r}
&=&\int\sqrt{-g}d\theta d\phi (-i)(S^{(0)\, *}\partial_{r}S^{(0)}
-\partial_{r}S^{(0)\, *}S^{(0)})\ , \nonumber\\
&=&r^2i(\partial_{r} R^{*}_{\ell}R_{\ell}-R^{*}_{\ell}\partial_{r}R_{\ell})
\ , \label{Bcurrent}
\end{eqnarray}
for scalar bosons and 
\begin{eqnarray}
J^{(k)}_{r}
&=&\int \sqrt{-g}d\theta d\phi 
\ i\bar{\psi}^{(k)}\, {\boldmath{\gamma}}\cdot
\frac{{\boldmath{x}}}{r}\, \psi^{(k)}\, \nonumber\\
&=&r^2 
i(\partial_{r} R^{*}_{\ell}R_{\ell}-R^{*}_{\ell}\partial_{r}R_{\ell})
\ , \label{Fcurrent}
\end{eqnarray}
for spinors with $k=1,+$ or $-$. 
Therefore we have obtained the current relation between scalar boson 
and spinor (spin up, spin parallel or spin antiparallel) as 
\begin{eqnarray}
J^{(0)}_{r}=J^{(1)}_{r}=J^{(+)}_{r}=J^{(-)}_{r}\ .
\label{BFcurrent}
\end{eqnarray}

\subsection{Super-radiance puzzle in Kerr black hole geometry}

In order to solve the super-radiance puzzle 
applying the Bargmann-Wigner formulation, 
we study the scattering problems for fermionic and bosonic fields 
in Kerr black hole geometry, of which metric is represented 
in Boyer-Lindquist coordinates \cite{Boyer1967}: 
\begin{eqnarray}
ds^2&=&\frac{\Delta}{\Sigma}[dt-a\sin^{2}{\theta}d\phi]^2
+\frac{\Sigma}{\Delta}dr^2+\Sigma d\theta^2
+\frac{\sin^2{\theta}}{\Sigma}[(r^2+a^2)d\phi-adt]^2\ , \\
\Delta&=&r^2-2Mr+a^2\ , \ \Sigma=r^2+a^2\cos^2{\theta}\ ,
\end{eqnarray}
where $M,a$ denote the mass and angular momentum of the Kerr black hole 
respectively.  

We first consider the scattering problem for the spin 0 scalar filed 
in Kerr geometry.   
The scalar field can be written in the polar coordinate system as
\begin{eqnarray}
S(x)=R(r)Y(\theta\, , \phi)\, {\exp}{(-i\omega t)}\ ,
\end{eqnarray}
where $R(r)$ and $Y(\theta\, , \phi)$ denote the radial and  
angular wave functions respectively, which obey the equations:
\bea
(\frac{1}{\sin{\theta}} 
\partial_{\theta}\sin{\theta}\partial_{\theta}
-(a\omega\sin{\theta}-\frac{m}{\sin{\theta}})^2-\mu^2a^2\cos^2{\theta} 
+\lambda)Y(\theta, \phi)&=&0 \ , \\
(\partial_{r}\Delta\partial_{r}+\frac{((r^2+a^2)\omega-am)^2}{\Delta}
-\mu^2r^2 -\lambda)R(r)&=&0 \ , 
\eea
where $\mu$ and $\lambda$ 
denote the mass of particle and separation parameter. 

To study the behavior of radial wave function 
near infinity and event horizon, a new radial coordinate $r^*$ 
is introduced: 
\begin{eqnarray}
\frac{dr^*}{dr}=\frac{r^2+a^2}{r^2-2M r+a^2} \ .
\end{eqnarray}
Using the new coordinate, 
radial field solutions in Kerr geometry become free waves 
near the spatial infinity 
$r\rightarrow \infty \ (r^{*}\rightarrow {\infty})$  
and event horizon 
$r\rightarrow r_{H}=M+\sqrt{M^2-a^2} \ (r^{*}\rightarrow {-\infty})$: 
\begin{eqnarray}
\sqrt{r^2+a^2}\,R(r) \sim 
\left\{
\begin{array}{cc}
{\omega}^{-1/2}(A^{(\rm inc)}_{B}\exp{(-ip_{\infty}r^*)}
+A^{(\rm ref)}_{B}\exp{(ip_{\infty}r^*)}) , & (r\rightarrow \infty)\\
{\mid p_{H}\mid}^{-1/2} 
A^{(\rm trans)}_{B}\exp{(-ip_{H}r^*)} , & (r\rightarrow r_{H})
\end{array}
\right. \ , 
\end{eqnarray}
where $A^{(\rm inc)}_{B}$, $A^{(\rm ref)}_{B}$ and $A^{(\rm trans)}_{B}$ 
denote scalar boson amplitudes of incident wave, reflected and transmitted 
waves respectively. 
Bosonic normalization factors are included in front of each amplitude.  
Momenta near the infinity and event horizon are denoted by 
\begin{eqnarray}
p_{\infty}=\sqrt{\omega^2-\mu^2}\ \ {\mbox{and }} \ \ 
p_{H}=\omega-\Omega_{H}m
\end{eqnarray}
which particle mass $\mu$, azimuthal momentum $m$ and 
angular velocity $\Omega_{H}=a/(r_{H}^2+a^2)$ respectively. 
The boson normalization factor is assumed to be included 
in the radial function $R(r)$ and consequently in the boson 
amplitudes $A^{(\rm inc)}_{B}\,, A^{(\rm ref)}_{B}$ and $A^{(\rm trans)}_{B}$.

Next we consider the scattering problem for the spin 1/2 spinor filed 
in Kerr geometry.
We can write the spinor wave function in the form:
\begin{eqnarray}
\psi(x)=\frac{1}{(\Delta\Sigma)^{1/4}}
\left(
\begin{array}{c}
F(r)\mathcal{Y}(\theta,\phi)\\
iG(r)\mathcal{Y}'(\theta,\phi)
\end{array}
\right)\ ,
\end{eqnarray}
where $\mathcal{Y}(\theta,\phi)\, , \mathcal{Y}'(\theta,\phi)$ 
stand for normalized spin-angular functions. 
Asymptotic radial component 
solutions for infinity and near event horizon are obtained 
\begin{eqnarray}
F(r) &\sim& 
\sqrt{\frac{\omega+\mu}{\omega}}(
A^{(\rm inc)}_{F}\exp{(-ip_{\infty}r^*)}
+A^{(\rm ref)}_{F}\exp{(ip_{\infty}r^*)}) \ , \\
G(r) &\sim& \frac{1}{(\omega+\mu)}\frac{d}{dr^*}F(r) \ , 
\end{eqnarray}
for infinity  $(r\rightarrow \infty)$ and 
\begin{eqnarray}
F(r) &\sim& 
A^{(\rm trans)}_{F}\exp{(-ip_{H}r^*)} \ , \\
G(r) &\sim& iF(r) \ ,
\end{eqnarray}
for near event horizon $(r\rightarrow r_{H})$, 
where $A^{(\rm inc)}_{F}$, $A^{(\rm ref)}_{F}$ and $A^{(\rm trans)}_{F}$ 
denote fermionic amplitudes of incident wave, reflected and transmitted 
waves respectively. Fermionic normalization factors are included in 
radial amplitudes as $\sqrt{(\omega+\mu)/\omega}$ 
for infinity and $1$ for near horizon because particles behave 
as massless ones.

We can derive the current conservation relation of the radial component 
between the infinity and event horizon 
\begin{eqnarray}
\frac{p_{\infty}}{\omega}
 (\mid A^{(\rm inc)}_{B}\mid^2-\mid A^{(\rm ref)}_{B}\mid^2)
= \frac{p_{H}}{\mid p_{H}\mid} \mid A^{(\rm trans)}_{B}\mid^2 \ , \label{CCR1}
\end{eqnarray}
for scalar bosons and 
\begin{eqnarray}
\frac{p_{\infty}}{\omega} (\mid A^{(\rm inc)}_{F}\mid^2
-\mid A^{(\rm ref)}_{F}\mid^2)
= \mid A^{(\rm trans)}_{F}\mid^2 \ , \label{CCR2}
\end{eqnarray}
for spinors.
The factor in front of $\mid A^{(\rm trans)}_{B}\mid$ 
is different of the factor in front of $\mid A^{(\rm trans)}_{F}\mid$, 
which is consistent for the massless neutrino cases.

In order to relate the bosonic and fermionic current, 
we apply the Bargmann-Wigner formulation at radial infinity, which is 
considered as free flat spacetime. We obtain the relation of conserved current 
between scalar boson and spinor at infinity region $(r\rightarrow \infty )$  
using equation (\ref{BFcurrent}) as 
\begin{eqnarray}
\mid A^{(\rm inc)}_{B}\mid^2-\mid A^{(\rm ref)}_{B}\mid^2
=
\mid A^{(\rm inc)}_{F}\mid^2-\mid A^{(\rm ref)}_{F}\mid^2 \ . \label{CCR3}
\end{eqnarray}
Combining three conserved current relations in equations (\ref{CCR1})
-(\ref{CCR3}), 
we get the conclusion that the momentum near the event horizon is 
positive:   
\begin{eqnarray}
0< p_{H}=\omega-\Omega_{H}m \ .
\end{eqnarray} 
This condition means that bosonic superradiance 
does not occur for $0< \omega$. 

It is worthwhile to note that bosonic and fermionic superradiance 
can occur for $\omega< 0$ with $0< \omega-\Omega_{H}m$. 

These conclusion do not change by the choice of normalization factors 
because they are positive definite.  

\section{Summary}
\setcounter{equation}{0}

We have studied the super-radiance puzzle between bosons and fermions 
in rotating black hole geometry using 
the generalized Bargmann-Wigner formulation. 

\begin{itemize}
\item
We have obtained the direct wave function relation between scalar bosons 
and spinors via bi-spinor fields under the extended Bargmann-Wigner formulation , established in the paper.  
\item
The superradiance phenomena for scalar bosons of the type 
$0<\omega$ and $\omega-m\Omega_{H}<0$ are shown not to occur 
in Kerr metric as those for fermion cases. 
\end{itemize}

Some discussions are added.
We have established the Bargmann-Wigner formulation for 
non-zero particle mass cases. 
We can take the massless limit after intermediate calculations because  
the mass is used as the regularization of the theory.
We expect that the superradiance problem of 
vector and tensor bosons will be solved rigorously 
as in the scalar boson case.

\section*{Acknowledgements}

I would like to give my thanks to Professor Kazuyasu Shigemoto 
for his discussions and careful reading to this manuscript. 

\section*{Appendix: Gamma matrices in flat and curved spacetime}
\setcounter{equation}{0}
Gamma matrix notation used in this paper is given in flat 
Minkowski spacetime:
\begin{eqnarray*}
  &&\gamma^{0}=\left(\begin{array}{cc}  
  I & 0\\
  0 & -I
  \end{array} \right),  \quad
  \gamma^{a}=\left(\begin{array}{cc}  
  0 & -i\sigma_{a}\\
  i\sigma_{a} & 0
  \end{array} \right),  \quad (a=1\sim 3)
\end{eqnarray*}
where $\sigma_{a} $ stand for Pauli matrices. 
Gamma matrices in curved spacetime are connected by the vierbeins as
$\gamma^{\mu}=b_{i}^{\mu}\gamma^{i}$, where Latin suffix $i=0\sim 3$ are for 
flat spacetime and Greek suffix $\mu$ are for curved spacetime. 
From the vierbein condition $\mathcal{D}_{\mu}b_{\nu}^{i}=0 $ in equation 
(\ref{condition}), 
differential equations for gamma matrices in curved spacetime are obtained:
\begin{eqnarray*}
\partial_{\mu}\gamma^{\nu}
=-\Omega_{\mu}\gamma^{\nu}+\gamma^{\nu}\Omega_{\mu}
-\Gamma^{\nu}_{\lambda \, \mu}\gamma^{\lambda}
=-\omega_{ij,\mu}\gamma^{i}b^{j\nu}
-\Gamma^{\nu}_{\lambda \, \mu}\gamma^{\lambda}\ ,
\end{eqnarray*}
where the connection matrix $\Omega^{\nu}_{\lambda \, \mu}$ and
the spin connection $\omega^{ij}_{,\mu}$ 
are defined in equations 
(\ref{Omega}) and (\ref{omega}) respectively.



\begin{thebibliography}{99}
\bibitem{Press1972}
W. H. Press and S. A. Teukolsky, Nature {\bf 238}, 211 (1972).
\bibitem{Chandrasekhar1983}
S. Chandrasekhar, {\it The Mathematical Theory of Black Holes}, 
Claredon Press (1983).
\bibitem{Cardoso2004}
V. Cardoso, O. J. C. Dias, J. P. S. Lemos and S. Yoshida, 
Phys. Rev. {\bf D70}, 044039 (2004).
\bibitem{Kodama2008}
H. Kodama, Prog. Theor. Phys. Suppliment {\bf 172}, 11 (2008). 
\bibitem{Teukolsky1972-1974}
S. A. Teukolsky, Phys. Rev. Lett. {\bf 29} 1114 (1972); 
S. A. Teukolsky and W. H. Press, Astrophys. J. {\bf 193} 443 (1974). 
\bibitem{Takasugi1997}
S. Mano and E. Takasugi, Progress of Theoretical Physics, 
{\bf 97}, 213 (1997).
\bibitem{Misner1972}
C. W. Misner, Phys. Rev. Lett. {\bf 28}, 993 (1972).
\bibitem{Detweiler1980}
S. Detweiler, Phys. Rev. {\bf D22}, 2323 (1980).
\bibitem{Mukohyama2000}
S. Mukohyama, Phys. Rev. {\bf D61}, 124021 (2000).
\bibitem{Maeda1976}
K. Maeda, Prog. Theor. Phys. {\bf 55}, 1677 (1976).
\bibitem{Unruh1974}
W. G. Unruh, Phys. Rev. {\bf D10}, 3194 (1974).
\bibitem{Kenmoku2008}
M. Kuwata, M. Kenmoku and K. Shigemoto, Class. Quantum Grav. {\bf 25}, 
145016 (2008). 
\bibitem{Kenmoku2008-2}
M. Kuwata, M. Kenmoku and K. Shigemoto, Prog. Theor. Phys. 
{\bf 119}. 939 (2008).
\bibitem{Bargmann1948}
V. Bargmann and E. P. Wigner, Proc. Nat. Acad. Sci. (USA) 
{\bf 34}, 211 (1948). 
\bibitem{Lurie}
As a text book, see for example, 
D. Lurie, {\it{Particles and Fields}}, Interscience Publications (1968).
\bibitem{Brill1957}
D. R. Brill and J. A. Wheeler, Rev. Mod, Phys. {\bf 29}, 465 (1957).
\bibitem{Weinberg}
As a text book, see for example, 
S. Weiberg, {\it{The Quantum Theory of Fields vol. III}}, 
Cambridge Univ. Press (2000). 
\bibitem{Boyer1967}
R. H. Boyer and R. W. Lindquist, J. Math. Phys. {\bf 8}, 265 (1967). 
\bibitem{Sakurai1967}
As a text book, see for example, 
J. J. Sakurai, {\it Advanced Quantum Mechanics}, Addison-Wesley Publishing 
Company (1967).

\end{thebibliography}
\end{document}